\documentclass[%
 reprint, 
groupedaddress,
 preprintnumbers,
 showkeys,
 amsmath,amssymb,
 aps,
]{revtex4-2}
\usepackage{xcolor}
\definecolor{darkblue}{RGB}{46,48,147}
\usepackage{url}

\usepackage{breakurl}
\usepackage[colorlinks=true,
            linkcolor=darkblue,
            urlcolor=darkblue,
            citecolor=darkblue,
            breaklinks=true]{hyperref}
\usepackage{xspace}
\usepackage{graphicx}
\usepackage{dcolumn}
\usepackage{bm}
\usepackage[normalem]{ulem}  

\graphicspath{{figures/}}

\definecolor{darkbluegrey}{rgb}{0.45, 0.43, 0.66}
\definecolor{vermillion}{rgb}{0.86, 0.18, 0.01}

\newcommand{\hlsfml}{\textsc{hls4ml}\xspace}

\newcommand{\BVIMIN}{\texttt{B:VIMIN}\xspace}

\newcommand{\BVIMAX}{\texttt{B:VIMAX}\xspace}
\newcommand{\BIMINER}{\texttt{B:IMINER}\xspace}
\newcommand{\BLINFRQ}{\texttt{B:LINFRQ}\xspace}
\newcommand{\BVIPHAS}{\texttt{B:VIPHAS}\xspace}
\newcommand{\IIB}{\texttt{I:IB}\xspace}
\newcommand{\IMDATFORTY}{\texttt{I:MDAT40}\xspace}
\newcommand{\IMXIB}{\texttt{I:MXIB}\xspace}

\providecommand{\NA}{\ensuremath{\cdots}\xspace}

\usepackage{scalerel}
\usepackage{tikz}
\usetikzlibrary{svg.path}

\definecolor{orcidlogocol}{HTML}{A6CE39}
\tikzset{
  orcidlogo/.pic={
    \fill[orcidlogocol] svg{M256,128c0,70.7-57.3,128-128,128C57.3,256,0,198.7,0,128C0,57.3,57.3,0,128,0C198.7,0,256,57.3,256,128z};
    \fill[white] svg{M86.3,186.2H70.9V79.1h15.4v48.4V186.2z}
                 svg{M108.9,79.1h41.6c39.6,0,57,28.3,57,53.6c0,27.5-21.5,53.6-56.8,53.6h-41.8V79.1z M124.3,172.4h24.5c34.9,0,42.9-26.5,42.9-39.7c0-21.5-13.7-39.7-43.7-39.7h-23.7V172.4z}
                 svg{M88.7,56.8c0,5.5-4.5,10.1-10.1,10.1c-5.6,0-10.1-4.6-10.1-10.1c0-5.6,4.5-10.1,10.1-10.1C84.2,46.7,88.7,51.3,88.7,56.8z};
  }
}

\newcommand\orcidicon[1]{\href{https://orcid.org/#1}{\mbox{\scalerel*{
\begin{tikzpicture}[yscale=-1,transform shape]
\pic{orcidlogo};
\end{tikzpicture}
}{0}}}}

\begin{document}

\preprint{FERMILAB-PUB-20-565-AD-E-QIS-SCD}

\title{Real-time Artificial Intelligence for Accelerator Control: A Study at the Fermilab Booster}

\author{Jason~St.~John\,\orcidicon{0000-0001-8110-4108}}
\email{stjohn@fnal.gov} 
\author{Christian~Herwig\,\orcidicon{0000-0002-4280-6382}}
\author{Diana~Kafkes\,\orcidicon{0000-0002-1716-463X}s}
\author{Jovan Mitrevski\,\orcidicon{0000-0001-9098-0513}}
\author{William~A.~Pellico\,\orcidicon{0000-0002-5009-8451}}
\author{Gabriel~N.~Perdue\,\orcidicon{0000-0001-6785-8720}}
\author{Andres~Quintero-Parra\,\orcidicon{0000-0001-6009-4959}}
\author{Brian~A.~Schupbach\,\orcidicon{0000-0001-7752-1451}}
\author{Kiyomi~Seiya\,\orcidicon{0000-0002-5057-6943}}
\author{Nhan~Tran\,\orcidicon{0000-0002-8440-6854}}
\affiliation{Fermi National Accelerator Laboratory, Batavia, Illinois 60510, USA}

\author{Malachi~Schram\,\orcidicon{0000-0002-3475-2871}}
\affiliation{Thomas Jefferson National Accelerator Laboratory, Newport News, VA 23606, USA}

\author{Javier~M.~Duarte\,\orcidicon{0000-0002-5076-7096}}
\affiliation{University of California San Diego, La Jolla, California 92093, USA}

\author{Yunzhi~Huang\,\orcidicon{0000-0002-5303-1658}}
\affiliation{Pacific Northwest National Laboratory, Richland, Washington 99352, USA}

\author{Rachael Keller}
\affiliation{Department of Applied Physics and Applied Mathematics, Columbia University, New York, New York 10027, USA\vspace{\baselineskip}}

\received{5 January 2021}
\accepted{16 August 2021}
\published{18 October 2021}


\begin{abstract}
We describe a method for precisely regulating the Gradient Magnet Power Supply (GMPS) at the Fermilab Booster accelerator complex using a neural network trained via reinforcement learning.
We demonstrate preliminary results by training a surrogate machine-learning model on real accelerator data to emulate the GMPS, and using this surrogate model in turn to train the neural network for its regulation task.
We additionally show how the neural networks to be deployed for control purposes may be compiled to execute on field-programmable gate arrays (FPGAs), and show the first machine-learning based control algorithm implemented on an FPGA for controls at the Fermilab accelerator complex.
As there are no surprise latencies on an FPGA, this capability is important for operational stability in complicated environments such as an accelerator facility.\\\\
  DOI: \href{https://doi.org/10.1103/PhysRevAccelBeams.24.104601}{10.1103/PhysRevAccelBeams.24.104601}
\end{abstract}

\maketitle


\section{Introduction}
\label{sec:intro}

Particle accelerators are among the most complex engineering systems in the world.
They are crucially important to the study of the elementary constituents of matter and the forces governing their interactions.
Tuning and controlling accelerators is challenging and time consuming, but even marginal improvements can translate very efficiently into improved scientific yield for an experimental particle physics program, where integrated accelerator run time imposes a substantial cost.

To date, the most common approach to accelerator systems control has largely consisted of hand tuning by experts, and has been guided by physical principles whenever possible.
However, accelerator physics is complex and highly nonlinear.
While we may model accelerator beams with impressive and improving precision~\cite{synergia}, building fully comprehensive Monte Carlo-based models of entire facilities is challenging, often leaving optimization and control to the intuition of experienced experts.
Even though this process has been successful to date, it is arduous and likely contains hidden inefficiencies.
Recently, deep learning~\cite{LeCun2015,Goodfellow-et-al-2016,Carleo:2019ptp}, the training of neural networks consisting of many hidden layers, has proven itself useful for complex control problems~\cite{deepcool,wiredcars,microsoftai,mitfanuc}.
Notably, processes within accelerator complexes occur between microsecond and millisecond timescales, much faster than human operators can react. 

In this study, we present a real-time artificial intelligence (AI) control system for precisely regulating an important subsystem of the Fermilab Booster accelerator complex~\cite{booster}. 
Our ultimate goal is to achieve a tenfold improvement in precision for the regulation system we describe in this paper.
To realize this system, we use machine learning (ML) for two key features: to develop \textit{surrogate models}~\cite{surrogate} that reproduce the behaviors of the real-world system, and to train an \textit{online agent} to take control actions in the system. 
The online agent is developed using reinforcement learning (RL)~\cite{sutton_barto_rl2018,Fran_ois_Lavet_2018}, a framework in which an artificial agent learns by interacting with its environment.  
This online agent will ultimately control the actual accelerator system.
The surrogate model provides an initial ``safe'' environment in which to train and evaluate control algorithms, and does so at higher rates and in a more controlled fashion than real-time training could allow.
In this work, to demonstrate the complete methodology, we show how the control algorithms perform within the surrogate model.

Additionally, we propose a scheme for the first ML control system implemented in FPGA firmware at the Fermilab accelerator complex. 
Our approach utilizes field-programmable gate arrays (FPGAs) for real-time responses and includes full integration into the Booster controls system for high-speed data ingestion.
On-board implementation, with dedicated hardware, is important for operational stability and for reliable low-latency response times.
In advance of commissioning the full system during online operations, we present important results from running a pretrained, static RL model in FPGA test benches.

Fully realizing this system will inevitably create an important and versatile set of tools that could find application in many areas in accelerator controls and monitoring, thereby paving the way for more ambitious control schemes.
These tools may allow for more complex agents to work in tandem to manage ever-larger portions of the accelerator complex.
While open questions remain as to whether data-intensive RL algorithms can be effectively used in this context without an independent, high-quality simulator, the results presented in this paper are encouraging.

In Sec.~\ref{sec:prevwork} we briefly highlight some previous publications at the intersection of machine learning and accelerator control.
Section~\ref{sec:complex} describes the Fermilab Booster accelerator and surrounding complex in sufficient detail to understand the specific controls problem we address.
In Sec.~\ref{sec:dataset}, we describe the accelerator data used to train a surrogate model, as well as the data processing used to support our ML workflow.
Next, in Sec.~\ref{sec:methods} we describe the ML algorithms for the surrogate model and candidate agents.
Then, in Sec.~\ref{sec:implfast} we briefly detail the process of deploying ML control algorithms to an FPGA.
Section~\ref{sec:summary} concludes with a discussion of our plans for full RL-based control. 

\section{Previous work}
\label{sec:prevwork}

The field of AI for accelerators is quite active in the development of new applications.
AI algorithms are poised to play a salient role in accelerator control, tuning, diagnostics, and modeling.
Here we provide a brief overview with a focus on recent work.

Predictive diagnostics are important precursors for control networks because they enable modeling of accelerator functions to a high degree of precision.
The authors of Ref.~\cite{PhysRevAccelBeams.21.112802} use ML-based diagnostics to predict the longitudinal phase space distribution in a particle accelerator. 
Reference~\cite{tennant2020superconducting} offers an example of AI applications for fault classification in superconducting radio frequency cavities. 
Furthermore, in groundbreaking conceptual studies~\cite{7454846,Edelen:2016jgh,Edelen:2016obn,Edelen:2017ewy,nncontrolforrapidswitchingneurips2017,yetanotherauraleefelpaper2017}, the authors built predictive networks for accelerator modeling that could serve as predecessors to control networks.
See also \cite{PhysRevSTAB.18.102801}.

In Ref.~\cite{PhysRevLett.121.044801} the authors use a neural network to demonstrate control of the longitudinal phase space of relativistic electron beams with very fine time resolution. 
However, as they and others have found, methods based on \emph{a priori} models face serious challenges in regard to managing complexity. 
A very recent study, found in Ref.~\cite{hirlaender2020modelfree}, explores some of the trade-offs between model-based and model-free training methods in an RL-based context and find some interesting advantages for each approach.
Here we seek to train, test, and deploy model-free RL control algorithms, but nevertheless must initiate that training using a surrogate model that provides a safe environment in which to evaluate algorithms before deployment. 
Surrogate models are an important ML-based tool for understanding other ML models (among other things) at accelerators where it is impractical to train from scratch using real accelerator hardware due to the risk of downtime and the absence of precisely repeatable history. 
The authors of Ref.~\cite{ogren2020surrogate} use neural network (NN) based surrogate models for the Compact Linear Collider (CLIC) final-focus system.
See also \cite{7454846,mlforoptcontrolxrayneurips2019,PhysRevAccelBeams.23.044601}.

Additionally, ML algorithms for tuning have been studied at the Linac Coherent Light Source (LCLS) at SLAC National Accelerator Laboratory, and at the Swiss Free Electron Laser (FEL). 
Bayesian optimization of the Swiss FEL is explored in  Ref.~\cite{DBLP:journals/corr/abs-1902-03229}.
Likewise, in Ref.~\cite{Duris_2020}, the authors use Bayesian optimization via Gaussian processes for fast tuning in response to the need to change beam configurations frequently.
Gaussian process models are sample efficient for producing accurate representations and uncertainties with limited data, but inefficient with respect to the number of samples in the dataset.
Furthermore, in Ref.~\cite{hanuka2020physicsinformed}, the authors extend the Gaussian process-based methods to employ physics-based simulations in cases where archival data is not available.  
The challenge addressed in this paper does not involve an explicit physical model for use by such a simulation.

Reference~\cite{alex2020advanced} summarizes a workshop that functions as a partial review of ML for accelerator controls and diagnostics and Ref.~\cite{arpaia2020machine} provides an overview of beam dynamics studies using AI at CERN. 
See also Ref.~\cite{Edelen:2018jid}.
In the most closely related previous work~\cite{cernrl}, the authors approach the controls problem of maximization of the CERN Low Energy Ion Ring multi-turn injected intensity in a similar way to how we approach rapidly cycling magnet power supply stabilization here. 
Some of the same authors very recently published Ref.~\cite{PhysRevAccelBeams.23.124801}, which studies continuous, model-free RL training algorithms at some different parts of the CERN accelerator complex.
Reference~\cite{boltz:icalepcs2019-tucpl06} similarly studies RL to control the RF system of the KIT Karlsruhe Research Accelerator (KARA) storage ring and improve microbunching instability. 
The authors of Refs.~\cite{electronics9050781,PhysRevAccelBeams.23.122802} deploy model-free RL training algorithms at the FERMI free-electron laser at Elettra with promising results for alignment in the first paper and for optimization and stabilization in the second.
As a proof of concept, the authors deploy a RL environment on an FPGA-ARM system for solving a classic cart-pole control problem~\cite{6313077}.

Fast NN inference has been explored on various edge devices, including FPGAs, for a variety of applications ranging from internet of things~\cite{banbury2020benchmarking} to high energy physics~\cite{CMSP2L1T}.
Surveys of some existing toolflows can be found in Refs.~\cite{2018arXiv180305900V,10.1145/3289185,Shawahna_2019,abdelouahab2018accelerating}, and include the \textsc{fpgaConvNet} library~\cite{venieris2017fpgaconvnet,venieris2017fpga,venieris2017fpl,venieris2016fccm}, \textsc{FP-DNN}~\cite{fpdnn}, \textsc{DNNWeaver}~\cite{dnnweaver:micro16}, \textsc{Caffeine}~\cite{caffeinatedFPGAs}, \textsc{Vitis AI}~\cite{vitisai}, the \textsc{FINN} project~\cite{blott2018finnr,FINN,finn_github}, \textsc{FixyNN} and \textsc{DeepFreeze}~\cite{whatmough2019fixynn,deepfreeze}, \hlsfml~\cite{hls4ml,hls4ml_github,DiGuglielmo:2020eqx,Summers:2020xiy,Coelho:2020zfu,Aarrestad:2021zos,Iiyama:2020wap,Heintz:2020soy,Fahim:2021cic}, and others~\cite{majumder2019flexible,7459526,hacene2018quantized,chang2020mix}.
Many of these toolflows are specific to Xilinx FPGAs. 
In this work, we build our implementation using the \hlsfml library as it has been extended to Intel FPGAs and allows for an easy exploration of the NN design space, as explained in Sec.~\ref{sec:implfast:fpga}.

Relative to the body of prior work on AI for accelerators, we offer new developments in the integration of realistic RL algorithms into embedded systems (FPGAs).
By embedding the algorithm in an FPGA, we may take advantage of the very stable, low-latency performance offered by that platform, which is critical for controls in a particle accelerator. 
Further, we indicate how this result may be extended through the use of a statistical ensemble of agent models, enabling increased learning rates, prediction stability, and robustness against potential mode collapse in individual agents. 
Given models of sufficiently small size, FPGA implementation of the models enables us to run an ensemble of models in parallel in order to optimize decision stability at no additional cost to overall latency. 

\section{Fermilab Booster Accelerator Complex}
\label{sec:complex}

\subsection{Accelerator Environment of the GMPS Regulator}
The Booster rapid-cycling synchrotron receives the 400\,MeV (kinetic energy) beam from the Fermilab Linear Accelerator (Linac) via charge-exchange (H$^{-}$ to H$^{+}$). 
This beam is accelerated to 8\,GeV by synchronously raising (``ramping'') the Booster accelerator cavities' frequency and the magnetic field of the combined-function bending and focusing electromagnets known as gradient magnets, which are powered by the gradient magnet power supply (GMPS)~\cite{rookie,Ryk:1974mu}. 
The beam is extracted at peak kinetic energy, after which the system is returned to the injection state. 
This complete cycle repeats, sinusoidally varying the GMPS magnet current between programmed current minimum and maximum at 15\,Hz.

\begin{figure}
\centering
\includegraphics[width=0.42\textwidth]{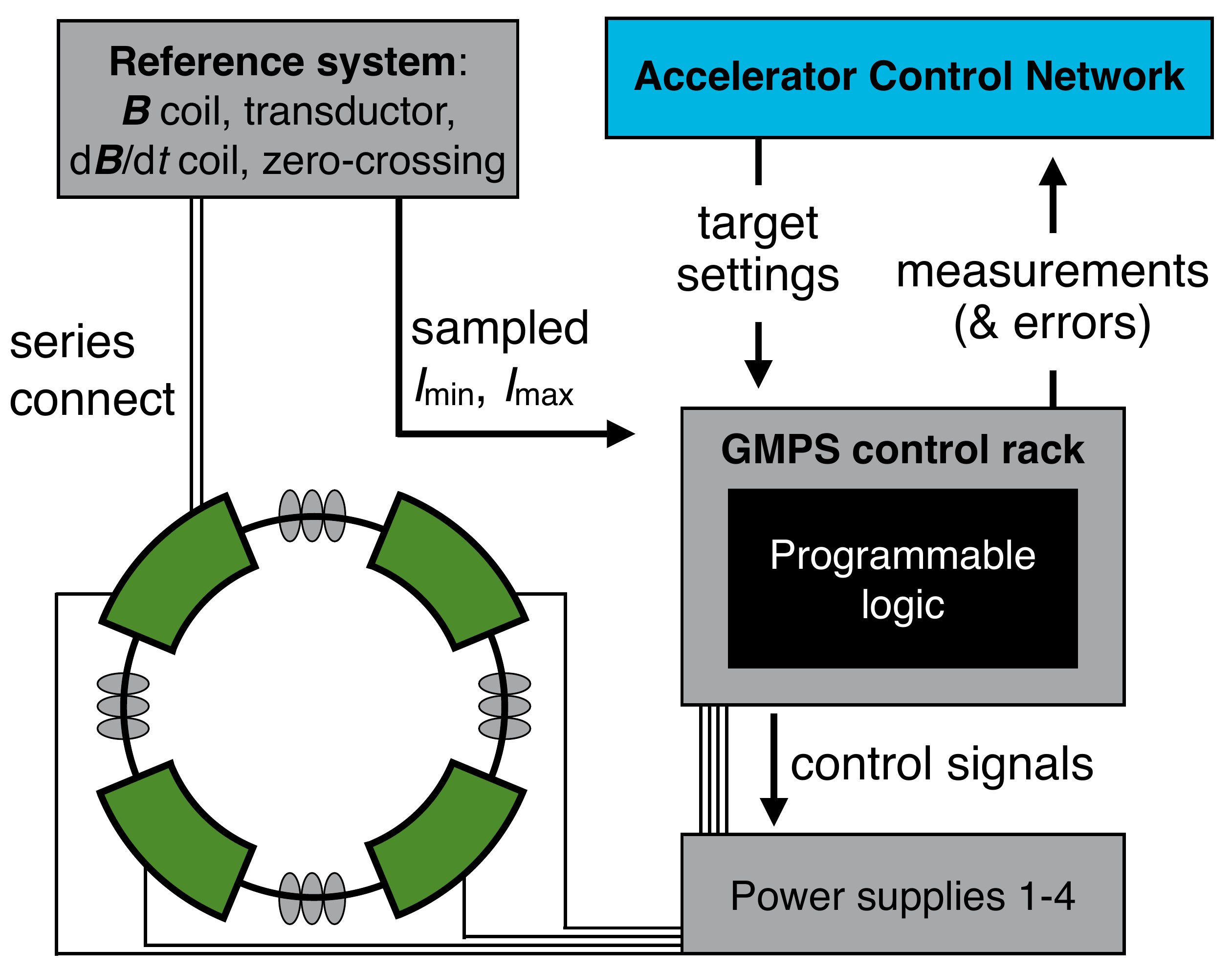}
\caption{Schematic view of the GMPS control environment.
The human operator specifies a target program via the Accelerator Control Network that is transmitted to the GMPS control board.
The FPGA-based control logic utilizes these settings together with readings from a reference magnet to prescribe a driving signal to the GMPS.
The effect of this prescribed signal on the bending magnets is measured by an in-series reference magnet, with sampled readings transmitted back to the GMPS control board.
Reference measurements and prescribed signals may be logged and transmitted over network for later analysis.}
\label{fig:schematic}
\end{figure}

Meanwhile, other nearby high-current, high-power electrical loads are varying in time, causing unwanted fluctuations of the actual GMPS electrical current, and thus fluctuations of the magnetic field in the Booster gradient magnets. 
The role of the GMPS regulator is to calculate and apply small compensating offsets in the GMPS driving signal, improving the agreement of the resulting minimum and maximum currents with their set points. 
The present GMPS regulator system is a proportional\textendash integral\textendash derivative (PID) controller~\cite{pid1,pid2}. 
Figure~\ref{fig:schematic} shows a schematic overview of the GMPS control environment.  
 
\begin{figure}
\centering
\includegraphics[width=0.48\textwidth]{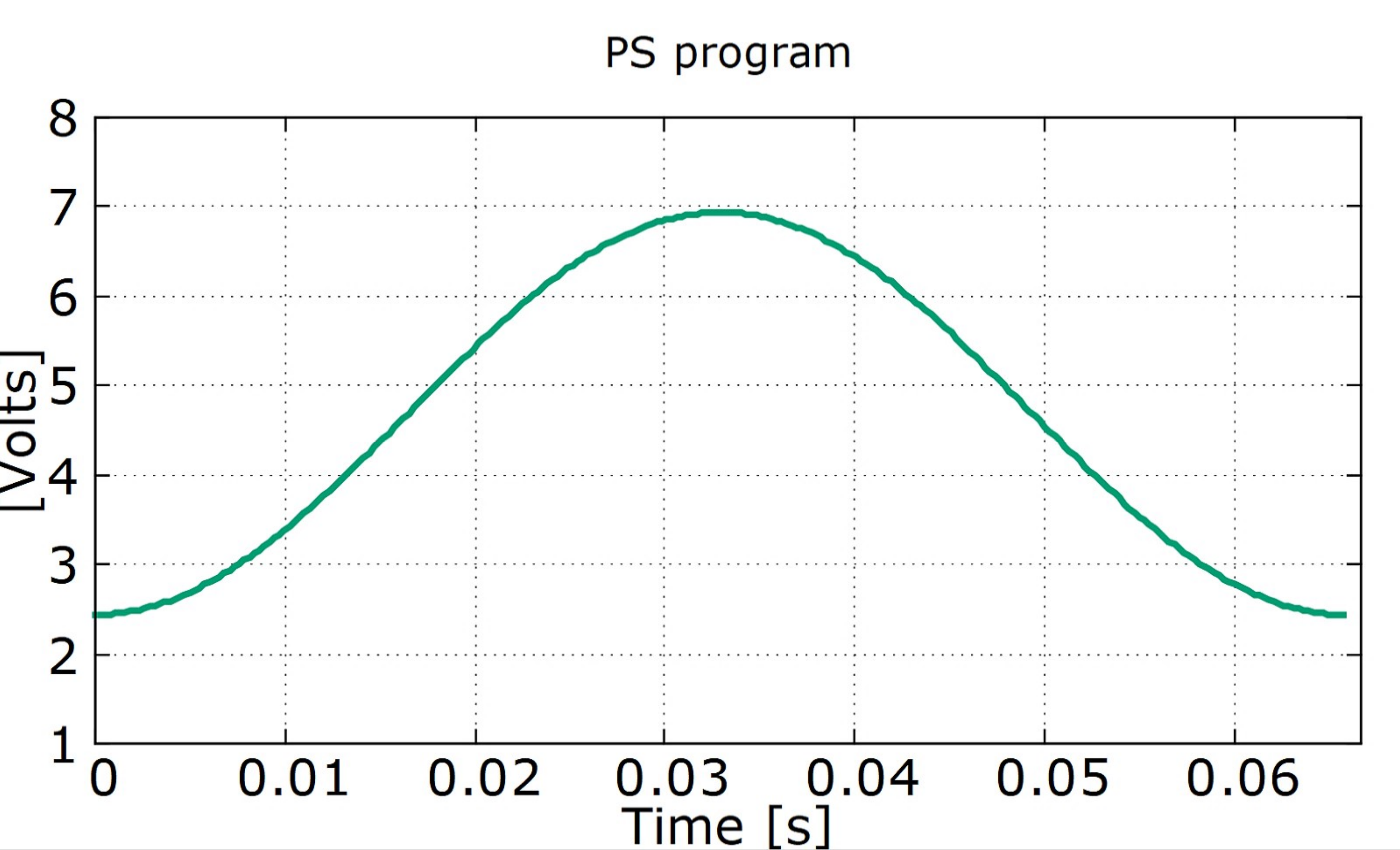}
\caption{A single 15~Hz cycle of the power supply program voltage, from one minimum to the next.}
\label{fig:outvolt}
\end{figure}

The power supplies that provide the combined DC and AC components of the desired gradient magnet current consist of a pair of three-phase series-connected silicon controlled rectifier~\cite{rookie} bridges fired at 720~Hz. 
An inductor-capacitor (LC) filter network at the output greatly reduces 720~Hz ripple, resulting in an output voltage that is proportional to the sinusoidal program provided by the GMPS regulator system. 
The series-connected electromagnet circuits and their cell capacitor banks are driven at resonance at 15~Hz and coupled via a distributed choke system for bypassing DC current. 
Figure~\ref{fig:outvolt} shows the power supply output voltage due to the cosine program.

For monitoring purposes, a special series-connected half-cell reference magnet located in the equipment gallery includes a pickup coil located between its poles to measure the time rate of change of the magnetic field $\dot{\vec{B}}$, which is an important input to the GMPS regulator. 
Powered with the other gradient magnets and housed in a low-radiation environment without charged particle beam passing through it, this reference magnet provides an accurate representation of the magnetic field under control throughout the accelerator.

Timing information derived from $\dot{\vec{B}}=0$ synchronizes the GMPS regulator system to the minimum and maximum values of the magnetic field and provides a transistor-transistor-logic (TTL) based 15~Hz master clock signal that drives the timing system for the GMPS regulator and indeed the rest of the accelerator complex. 
The high-frequency sampled measurements near the minimum and maximum values of the magnetic field are automatically fitted each cycle, and the finite impulse response (FIR) parameters are used as the primary feedback mechanism for the GMPS regulator system. 
Reducing the errors (the difference between the target and realized GMPS current especially at injection) of the GMPS system is of primary concern in the operational performance and efficiency of the Booster. 
The following Sections discuss the details of the present and proposed regulation systems.

\subsection{GMPS Regulation}
\label{sec:currreg}

The GMPS regulation system seeks to minimize the impact of disturbances due to environmental factors such as ambient temperature, nearby high-power pulsed RF systems, and ramping power supplies with inductive loads. 
Variations in the AC line frequency and amplitude are also significant sources of error, and are due in part to other particle accelerators in the complex changing currents in their own high-current electromagnets as part of their normal operations. 
Without regulation, the fitted minimum of the magnetic field may vary from the set point by as much as a few percent.

The existing regulator reduces GMPS regulation errors to roughly 0.1\% of the set value by implementing a proportional-integral-derivative (PID) control scheme.
Each cycle, the fitted minimum and maximum of the magnetic field reflect the combined influences of the set points, any compensation applied by the regulator for that cycle, and any new influence of other nearby electrical loads. 
Calculated estimates for the minimum and maximum values of the changing magnetic field of the previous 15~Hz cycle are used to adjust the power supply program to decrease the errors of the system.  
The calculation of the compensated minimum current \BVIMIN, used in the next cycle of the power supply program voltage, proceeds as follows at each time step $t$:
\begin{align}
\label{eqn:PIDregulator}
    \BIMINER(t) &= 10[I^\mathrm{min}_\mathrm{fit}(t) - I^\mathrm{min}_\mathrm{set}(t)]\,, \\
    \beta(t) &= \Gamma(t)\BIMINER(t) + \beta(t-1)\,,\\
    \BVIMIN(t) &= I^\mathrm{min}_\mathrm{set}(t) - \alpha(t)\BIMINER(t) - \beta(t)\,,
\end{align}
where $\BIMINER(t)$ and $\BVIMIN(t)$ are the corresponding control system readings at time $t$; $I^\mathrm{min}_\mathrm{fit}$ is the measured (fitted) current minimum of the present cycle; $I^\mathrm{min}_\mathrm{set}$ is the nominal set value of the current minimum; $\Gamma$ is the integral gain; and $\alpha$ is the proportional gain.  
The running parameter $\beta$ is effectively an integral of the error which rapidly downweights past measurements of the error. 
Typical values of the gain constants are $\Gamma = 7.535\times 10^{-5}$ and $\alpha=8.5\times 10^{-2}$, adjustable by system experts. 
\BIMINER is proportional to the error in the system, and for our purposes they are synonymous. 

\begin{figure}
\centering
\includegraphics[width=0.49\textwidth]{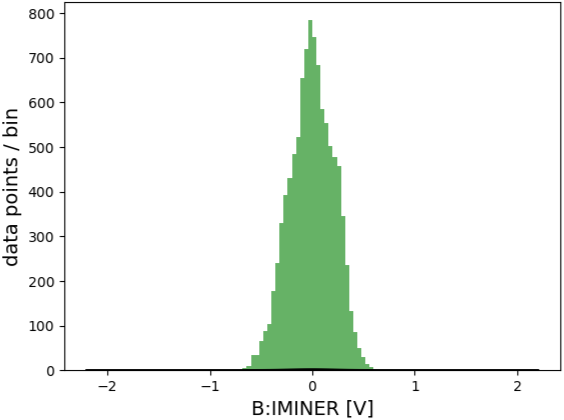}
\caption{Distribution of fractional measured error in the GMPS current at the minimum value of the magnet current, with the non-ML PID regulator discussed in the text.}
\label{fig:errdist}
\end{figure}

The environmental perturbations, discussed above, increase the distributed long-term steady-state errors of the GMPS system. 
A traditional PID regulation loop, given a sufficient amount of time and unchanging perturbations, will decrease the steady-state error of a system to zero. 
In reality, and within the timescale of the Booster beam cycle, the PID loop decreases the steady-state error to some nonzero error, typically of order 0.1\% for the cycle minimum. 
See Figure~\ref{fig:errdist} for a sample distribution of measured errors for the minimum value of the sinusoidally varying magnet current.
Efforts to decrease the steady-state error further by adjusting the closed-loop gains to be more responsive would come at the cost of reduced overall system stability. 
Therefore, a balance between the steady-state error and stability of the system has been struck. 

We seek to improve upon this regulation performance in order to decrease injection losses and improve Booster efficiency. 
An RL model, whose inputs include all of the most important outside influences on the GMPS system, can infer more accurate cycle-by-cycle compensations on an FPGA, reliably producing better regulation against those influences. 
Creating such an algorithm begins with collecting time-series data from the accelerator complex in operation. 

\section{Dataset} 
\label{sec:dataset}

We collected a dataset for Booster GMPS regulation to provide cycle-by-cycle time series of readings and settings from the most relevant devices available in the Fermilab control system~\cite{Cahill:2008zz}.
This data was drawn from the time series of a select subset of the roughly 200,000 entries that populate the device database of the accelerator control network~\cite{ACNET}. 
Data was sampled at 15~Hz for 54 devices that pertain to the regulation of the GMPS during two separate periods of time: Period 0 (June 3, 2019 to July 11, 2019) was ended by the annual Summer Shutdown and Maintenance. Period 1 (December 3, 2019 to April 13, 2020) ended when the accelerator operations were suspended in response to the COVID-19 pandemic.
In this paper, we use a subset of this dataset's devices and a single day---March 10, 2020---for development and demonstration, as detailed in Sec~\ref{sec:vbacmodel}.
For a more detailed overview, please see the corresponding Data Descriptor available online~\cite{BOOSTR:datasheet}. To our knowledge, this is one of the first well-documented datasets of accelerator device parameters made publicly available.
We strongly encourage our colleagues on other accelerator ML application projects to do likewise; reproducible results are the foundations of science.

In this accelerator complex data-logging nomenclature, device parameters with the \texttt{B:} prefix are related to the Booster, whereas device parameters beginning with \texttt{I:} are related to the Main Injector. Additionally, ``MDAT" denotes the accelerator (machine) data communication broadcast. 

\subsection{Variable Selection}
\label{subsec:devices}

We chose a subset of the 54 device time series available to facilitate initial studies of \BIMINER, the measure of regulation error at each GMPS cycle minimum.
These first studies (which are presented here) were conducted using five sets of time-series data suggested by accelerator domain experts to be the most important for regulating \BIMINER.
In addition to \BIMINER, these include \BLINFRQ, \BVIMIN, \IIB, and \IMDATFORTY.
Further optimization will be pursued prior to deployment of the production system.

Here, \BVIMIN is the compensating recommendation for the minimum value of the offset-sinusoidal GMPS current, issued by the GMPS regulator in order to reduce the magnitude of \BIMINER. 
\BLINFRQ is the measured offset from the expected 60~Hz line frequency powering the GMPS. 
\IIB and \IMDATFORTY provide measurements of the main injector bending dipole current at different points in the circuit and through different communication channels.

This expert-chosen set of just five parameters was used to train the surrogate model with the RL agent described in Sec.~\ref{sec:methods}, and to characterize RL agent models on an FPGA in Sec.~\ref{sec:implfast}. 

As a check on this selection, a Granger causality study ~\cite{granger} was performed using those variables correlated or anticorrelated with \BIMINER, with absolute Pearson correlation coefficient $|r| > 0.20$. 
Additionally, \BLINFRQ and \IIB were included in the study on the advice of system experts.
The Granger causality study, described here, allows us to explore the utility of the full set of logged signals for future studies, with minimal human bias.

For a given pair of concurrent time series---one potentially causal and the other responsive (one affects the other)---Granger causality does not prove pure causation.
Instead, Granger causality suggests the response variable could be better predicted using both time series jointly rather than using the response variable's self-history alone.
This test consists of creating and comparing two linear regression models: a self-model and a joint-model for both the response variable and the potentially causal variable, and calculating coefficients for each lag value (time difference) being tested, from one up to some predetermined maximum number of time offsets ~\cite{granger}. 
If at least one coefficient is not zero in the joint model, then the other variable is said to be ``Granger causal'' with respect to the response variable, at the lag value being tested.
We compared p-values to test statistical significance at each lag value up to 50 lags (approximately 3.33\,s) as well as looked at the difference between the Bayesian information criterion of the self and joint-models. 
As a result of these iterative calculations, we identified three additional variables---\BVIMAX, the compensated maximum GMPS current, \BVIPHAS, the GMPS ramp phase with respect to line voltage, and \IMXIB, the main injector dipole bend current---that will be considered in the next iteration of the surrogate model.

The expert-selected devices \BIMINER, \BLINFRQ, \BVIMIN, \IIB, and \IMDATFORTY were indeed a subset of the top eight ``causal'' variables identified through the causality study, boosting confidence in their utility. Table~\ref{tab:parameters} briefly summarizes the parameters of interest used in this iteration of the surrogate model.

\begin{table}[ht!]
\caption{Description of dataset parameters chosen by experts and later validated with a causality study. 
Here,``MI" means Main Injector, ``MDAT" means accelerator (machine) data communication, and device parameters that begin with \texttt{B} are related to the Booster, whereas device parameters that begin with \texttt{I} are related to the Main Injector.
}

\centering
\resizebox{\columnwidth}{!}{
\begin{ruledtabular}
\begin{tabular}{l|l}
Parameter & Details [Units] \\ \hline
\BIMINER & Setting-error discrepancy at injection [A] \\ 
\texttt{B:LINFRQ} & 60~Hz line frequency deviation [mHz] \\ 
\BVIMIN & Compensated minimum GMPS current [A] \\ 
\IIB & MI lower bend current [A] \\ 
\IMDATFORTY & MDAT measured MI current [A] \\ 
\end{tabular}
\end{ruledtabular}}
\label{tab:parameters}
\end{table}

\subsection{Data Processing for ML}
\label{subsec:pipeline}

Although the devices were configured to write out reading and setting data at 15~Hz, actual timestamp intervals varied from this nominal frequency, and timestamps were not well synchronized across devices.
Thus, for time alignment purposes, we made use of the recorded timestamps for \texttt{Event0C}, a broadcast accelerator control event which is synchronized to the fitted minimum of the periodically varying magnetic field in the gradient magnets described in Sec.~\ref{sec:complex}.
This control event serves as a logical choice of reference time for GMPS-related parameters.  
Using Apache Spark-based~\cite{10.1145/2934664} algorithms to distribute data processing in parallel, we first calculated the maximum interval between successive timestamps for each device across all 176 days (necessarily excluding the five-month gap between our two data-taking periods).
We then used the corresponding largest observed lag between recorded values within each device to place the upper limit on a look-back window from an \texttt{Event0c} timestamp, and took for every \texttt{Event0C} the most recent device datum, whose timestamp was within that window for that device.
For more details on the data pre-processing decisions made in the creation of this dataset please see our Data Descriptor~\cite{BOOSTR:datasheet}.

\section{Machine Learning Methods}
\label{sec:methods}
Machine Learning (ML) refers to the process by which we adjust the randomly initialized parameters of generic function approximators, termed ``models,'' so as to minimize an appropriately chosen loss function, or conversely to maximize a reward function.  
As used in ML, feed-forward neural network model architectures specify arrangements of ``nodes,'' usually co-evaluated in layers, where each node calculates the weighted sum of the inputs and a bias term, and outputs the value of a nonlinear ``activation function.''
In the simple multilayer perceptron (MLP) architecture, all nodes in each layer send copies of their output values to all nodes in the next layer. 
Layers not at the input or output are termed ``hidden'' layers.
There are useful variations on this simple layer stack architecture such as recurrent neural networks (RNN), wherein some outputs from a previous forward inference are taken as inputs. 
Our work makes use of both MLP and RNN architectures. 

For a given architecture with its activation functions, the weights and biases are the parameters being adjusted in the optimization or ``training.'' 
The ``learning rate'' sets the proportionality of parameter adjustment to gradients of improved performance (based on lower loss or higher reward), and the development of sophisticated optimization schemes is an active research area.

\begin{figure}
\centering
\includegraphics[width=0.42\textwidth]{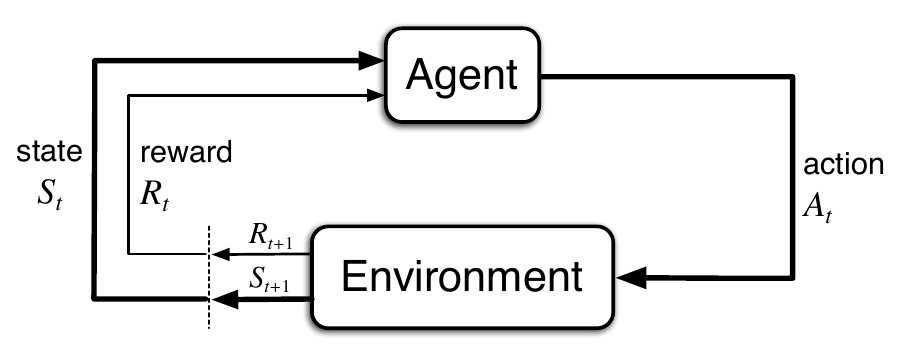}
\caption{The agent-environment interaction in a Markov decision process~\cite{sutton_barto_rl2018}. 
The agent executes a policy that selects an action $A_t$ given the current $S_t$, which results in a reward $R_{t+1}$ and a new state $S_{t+1}$ of the environment.}
\label{fig:rl-model}
\end{figure}

Reinforcement Learning (RL) is the subfield of AI aimed at optimizing of control or planning of complex tasks based on iterative feedback inputs from an environment, as explained below. 
The main components of RL are the environment and the agent, an ML model, as illustrated in Fig.~\ref{fig:rl-model}. 
RL trains an agent model over many time steps, and the resulting agent model may then be taken as fixed, to be deployed on new data. The agent model may be expected to perform similarly on new data as it did in training, provided the dynamics of the new data were well represented in the training data.
However RL extends naturally to continuous online learning, which would allow our regulator to adapt to changing environmental dynamics such as seasonality or new modes of accelerator complex operation, even though we would initially deploy a static or infrequently updated model out of prudence.
We set out to use RL to train an optimal regulation policy, which dictates an appropriate action that the GMPS regulator should take for any given state of the system.
Of course, as was highlighted in Sec \ref{sec:prevwork}, methods other than neural network-based RL have had success in control problems and practitioners should assess and compare these approaches as well.

The environment, usually formulated as a Markov decision process, is represented by a time-independent, discrete system with which the RL agent interacts (e.g. the accelerator complex).  
For the regulation of the GMPS current minimum, the environment includes the time-varying power demands and outside electrical influences for which the GMPS regulator makes compensating changes.
At each time step $t$, the environment takes in the control action $A_t$ determined by the RL agent based on the current state $S_t$, and provides the new system state $S_{t+1}$ (e.g. settings and measured quantities) along with an associated reward $R_{t+1}$.
Optimizing the agent's policy actions is defined to mean maximizing the long-term integrated reward, which is calculated over each fixed episode. 
In this study, the reward is calculated from the error in the minimum value of the GMPS current, \BIMINER:
\begin{equation}
R_t = -|\BIMINER(t)|\,.
\label{eqn:reward}
\end{equation}
The larger the magnitude of \BIMINER, the lower the reward.  
The possible actions $A_t$ we consider correspond to overriding the fixed setting of \BVIMIN, the lone control variable, with small, compensating adjustments.
This enables the agent's policy model to control \BIMINER without the PID regulator's input.

Recently, significant progress has been made in RL by combining it with advances in deep learning.
Deep learning models are well suited to representing complex policies for high-dimensional problems such as regulation in a dynamically variable environment.  
The deep $Q$-network (DQN)~\cite{dqn,mnih2013playing} approach, which we adopt for this study, involves using a deep neural network to learn the action-value function, or $Q$-value, and is usually deployed in environments that take discrete control actions.
The optimal policy can then be derived by choosing the action that maximizes the expected $Q$-value.

More formally, a policy $\pi$ is used by an agent to decide what actions $A_t = \pi(S_t)$ to take given a state $S_t$ at time $t$. 
An optimal policy $\pi^*$ maximizes the  $Q$-value, 
\begin{equation}
Q(S_t,A_t) = \sum_{t'=t}^{T} \mathbb E \left [ \gamma^{t'-t} R(S_{t'}, A_{t'}) | S_t, A_t \right ]\,,
\end{equation} 
where $\mathbb E$ is the expectation value operator, $R_t\equiv R(S_t, A_t)$ is the reward at time $t$, and $\gamma$ is the discount factor that de-emphasizes future rewards relative to present ones.
For this study, we used a value of $\gamma=0.85$.
The $Q$-value is the sum of the expected discounted rewards from the current time $t$ up to the horizon $T$.
In practice, the optimal action-value function $Q^*$ is not known a priori, but it can be approximated iteratively because it satisfies the Bellman equation~\cite{bellman},
\begin{equation}
Q^*(S_t,A_t) =  \mathbb E \left [R_t+\gamma\max_{A_{t+1}} Q^*(S_{t+1},A_{t+1})| S_t, A_t\right ]\,.
\end{equation}

In a DQN, the $Q$-value is approximated using a deep neural network, or policy model, with parameters $\theta$.  
In particular, the loss function at a time $t$ is given by the mean squared error (MSE) in the Bellman equation,
\begin{equation}
L_t(\theta_t) = \mathbb E\left [( y_t - Q(S_t, A_t; \theta_t))^2 \right]\,,
\end{equation}
where the (unknown) optimal target values are replaced by the approximate target values $y_t = R_t + \gamma\max_{A_{t+1}}Q(S_{t+1},A_{t+1}; \theta_t^{-})$ using parameters $\theta_t^-$ derived from previous iterations.

Continuous action space environments, such as the compensating adjustments of our GMPS current regulator, can adopt the DQN algorithm by discretizing the action space. 
To keep the DQN action space finite, we discretize the \emph{change} of control signal \BVIMIN using steps of just a few different sizes, including the option for zero-size change.  
By doing so, we explicitly limit the control signal variation between time steps, while helping to minimize the tuning and possible resonance of the overall surrogate-RL training loop.  
Other RL algorithms, such as deep deterministic policy gradient (DDPG)~\cite{ddpg}, proximal policy optimization~\cite{schulman2017proximal}, and twin delayed DDPG~\cite{fujimoto2018addressing} can provide continuous control suggestions which fit the GMPS regulator more naturally. 
However, the sample efficiency and tuning stability of these algorithms presents difficulties as discussed in various studies~\cite{Nair2018,HER}.  

The trial and error nature of RL training requires an environment that accommodates offline iteration. 
As mentioned in Sec.~\ref{sec:prevwork}, most real-world, complex systems cannot afford such an approach due to the risk of system failure. 
Therefore some level of offline training is essential.  
To facilitate offline training of a control agent, here we develop a surrogate model using the select collection of historical measurements described in Sec.~\ref{sec:dataset}. 
This surrogate captures the subset of the Booster accelerator complex necessary to model the dynamics of GMPS regulation in its environment.

In the following subsections, we discuss the workflow employed to develop and use an offline policy optimization, a prudent step prior to deploying the agent model in the real system.
Sec.~\ref{sec:vbacmodel} details the surrogate model that is developed to test the deep RL agent.  
Then, in Sec.~\ref{sec:rl}, we describe how the agent itself is trained.

\subsection{Virtual Accelerator Complex Model}
\label{sec:vbacmodel}

In order to train the RL policy model, we adopted a long short-term memory (LSTM)~\cite{lstm_paper} architecture for a surrogate of the GMPS regulator in its operating environment using data which describe the GMPS current and the Main Injector operation cycle. 
Its purpose is to predict the impact of changing the control variable \BVIMIN on \BIMINER over the course of the RL episodes.

An LSTM model is a specific type of recurrent neural network (RNN), which is appropriate for modeling sequential data such as our accelerator data. 
The key feature of an RNN is the network recursion, which enables it to describe the dynamic performance of systems over sequential time steps.
One difficulty of training RNNs is the vanishing gradient problem~\cite{pmlr-v28-pascanu13,279181}, which is a result of the gradient at a recursion step $k$ being dominated by the product of partial derivatives of hidden-state vectors $h$, $\prod_{j=k}^{R-1} \partial h_{j+1}/\partial h_j \propto w^{R-k-1}$, where $R$ is the number of recursions and $w$ is a recurring weight. 
If $w<1$, then this product tends to zero for large $R$.
The design of the LSTM includes an internal memory state, which effectively mitigates the vanishing gradient problem, and allows the network to retain a longer memory of past inputs~\cite{lstm_paper}.
Given the complexity of the GMPS regulation environment, we selected the LSTM architecture in order to capture the multiple frequency modalities observed in the data.

The details of the surrogate model's architecture are given in Table \ref{tab:model_pams}.
At each time step, the input to the Booster ML surrogate model is the most recent 150 time steps (equivalent to 10 seconds of data) from the five input variables selected in Sec.~\ref{subsec:devices}: \BVIMIN, \BIMINER, \BLINFRQ, \IIB, and \IMDATFORTY, and the output is a prediction for the values of \BIMINER, \BLINFRQ, and \BVIMIN at the next time step. 
While the latter two values are not included in the environment state propagated to the RL agent, the additional output structure helps to regularize the training process and validate performance.
The LSTM model uses this 150-step look-back window to ``recall'' the historical patterns of the time-series data, and thus achieve high accuracy in prediction.

The LSTM surrogate model was developed and implemented using the \textsc{Keras} library~\cite{keras}.
We used the Adam optimizer~\cite{adam} and a cost function of the mean squared error (MSE) of the predictions. 
The total number of data samples used for the analysis of the LSTM surrogate was 250,000 time steps from March 10, 2020, which we split into two non-overlapping data sets composed of 175,000 time steps for training and 75,000 time steps for testing. These data sets were then processed to allow 150 time step look-back, 1 time step look-forward as mentioned above.

\begin{figure}
\centering
\includegraphics[width=0.48\textwidth]{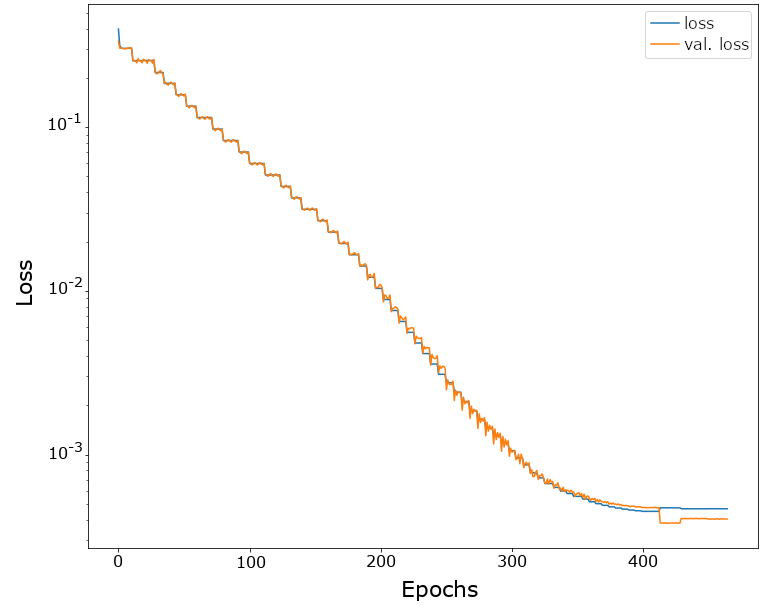}
\caption{Loss function as a function of the number of training epochs for the Booster LSTM surrogate model. 
The blue line gives the loss values for training sample and the orange line is the calculated loss using validation samples.}
\label{fig:surrogate-model-loss}
\end{figure}

The training samples were then further split using a K-fold cross-validation method: we defined five cross-validation folds that split the training and validation in a 80\%/20\% split.
This technique was used in order to estimate how the surrogate model is expected to perform in general as well as to monitor over-fitting.
While this sort of cross-validation was performed on the same segment of data in this implementation, we plan to cross-validate on different data samples when training the surrogate model in the future, at a larger scale.
The loss values from the validation sample were used to determine if the learning rate should be reduced or if the surrogate model had stopped learning, as shown in Figure~\ref{fig:surrogate-model-loss}.  
After more than 300 training epochs, the figure shows a bifurcation between the values of training loss and validation loss, suggesting some over-fitting.
Therefore, we used the values of model parameters prior to this bifurcation as the parameters of our surrogate model.
On separate test data, the loss value for this surrogate model was determined to be $9 \times 10^{-4}$ which is consistent with the training data set prior to the bifurcation.

\begin{table}[!t]
\label{tab:model_pams}
\renewcommand{\arraystretch}{1.3}
\caption{Fermilab Booster surrogate model, which learns to reproduce the environment in terms of the three time-series variables, one of which determines the reward as given in Eq.~\ref{eqn:reward}. 
The input LSTM layer receives five values, describing the current state \BIMINER, \BLINFRQ, \BVIMIN, \IIB, and \IMDATFORTY. The output layer is a prediction of \BIMINER, \BLINFRQ, \BVIMIN.}
\centering
\begin{ruledtabular}
\begin{tabular}{r|c|c|c|c}
Layer & Layer Type & Outputs & Activation & Parameters \\
\hline
1 & LSTM & 256 & $\tanh$ & 416,768 \\
2 & LSTM & 256 & $\tanh$ & 525,312 \\
3 & LSTM & 256 & $\tanh$ & 525,312 \\
4 & dense & 3 & linear & 771 \\
\hline
Total & \NA & \NA & \NA & 1,468,163 
\end{tabular}
\end{ruledtabular}
\end{table}

\begin{figure}
\centering
\includegraphics[width=0.5\textwidth]{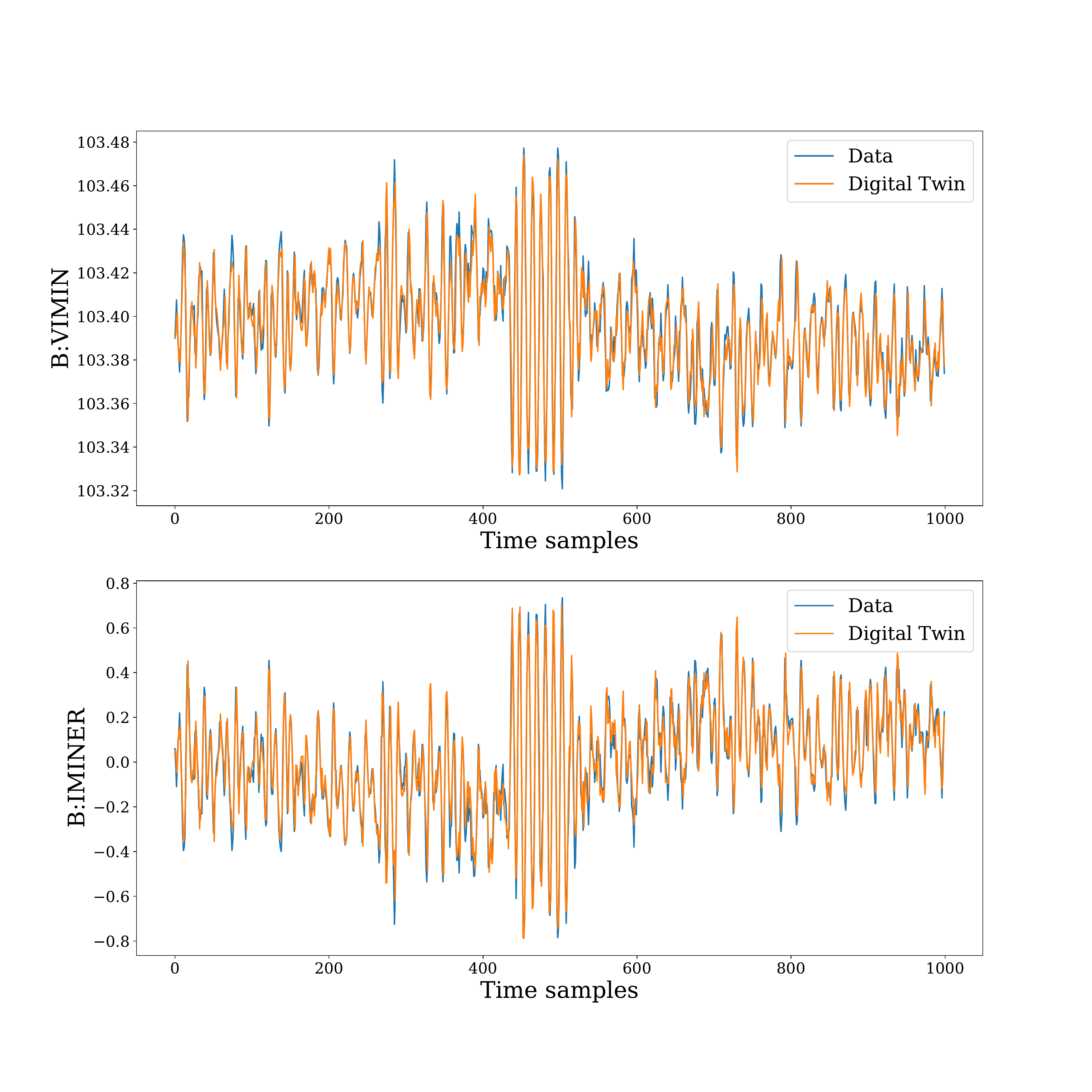}
\caption{Selected test data (blue) versus prediction values (orange) from the Booster LSTM surrogate model. New data is fed into the trained surrogate model at each time step.}
\label{fig:surrogate-model}
\end{figure}

Overlaid time series from the data and from LSTM predictions for a selected time window are shown in Fig.~\ref{fig:surrogate-model}.
Based on the great similarity of these results, the surrogate model was deemed adequate to use for initial training of the RL agent policy model.

\subsection{Reinforcement Learning for GMPS Control}
\label{sec:rl}
For this study, we formulated the problem as an episodic Markov decision process, where every episode contains 50 time steps. 
As in all $Q$-learning, the agent learns to maximize the reward within the time horizon of an episode. 
We developed our RL workflow based on a variant of DQN, the double DQN algorithm~\cite{dqn,NIPS2010_3964,vanhasselt2015deep}, using \textsc{Keras}~\cite{keras} to optimize \BVIMIN settings dynamically to minimize GMPS error \BIMINER.
The double DQN explicitly decouples the \emph{target model}, which is used to evaluate actions, from the \emph{policy model}, which is used to select actions, although they take the same form.

We used the \textsc{OpenAI Gym} package~\cite{openai_gym} to develop the environment that serves as a wrapper around the virtual accelerator complex model described above in Sec.~\ref{sec:vbacmodel} to interact with the RL agent.
The observation state space is simply the aforementioned five variables in the surrogate model section above, shown to causally relate to the measure of regulation error, \BIMINER. 
The action state space only contains one free parameter of control: adjustments to \BVIMIN.  
The seven discrete control options relative to the previous \BVIMIN are 0 (no change), $\pm0.0001$, $\pm0.005$, and $\pm0.001$. 
The choice of these values was based on the actual distribution of the changes in \BVIMIN observed in the data. 

At the start of each episode, 150 time steps from the data are used to initialize the system state, as is required for the environment surrogate model.
The 150th step defines the system state used by the agent. 
For each step thereafter, the agent provides a new action by specifying a change to \BVIMIN, and the system state is updated. 
The new system state is then used by the surrogate model to predict the resulting value of \BIMINER.
After the prediction the system state is incremented to the next time step. 
The current state, reward, and status for each step is passed to the agent to be used for training the DQN policy model.
RL algorithms learn from the reward provided by the environment, which in this study is given by Eqn.~\ref{eqn:reward}. 

During training, event samples are placed into a buffer before calculating the loss.
This \emph{memory buffer} is sampled randomly in a process called experience replay~\cite{dqn} in order to remove instabilities found to arise from training on time-ordered samples.
Once the memory buffer has sufficient experiences (32 experiences for this study) the active policy model begins training and continuously updating.
We use the $\epsilon$-greedy~\cite{Watkins:1989} method to control the agent's trade-off between exploration (random choice of action) and exploitation (deterministic action dictated by the current policy), in which the optimal action according to the current policy is chosen with probability $1-\epsilon$, while a random action is selected with probability $\epsilon$.
At the beginning of the training session we set $\epsilon=1$ with a decay factor of $0.9995$, applied multiplicatively whenever an exploration action is selected, until a minimum value of $\epsilon=0.0025$ is reached.
For this study, we use a multilayer perceptron (MLP) as the policy model (and target model) architecture, and rectified linear unit (ReLU) activation functions~\cite{relu}, as summarized in Table~\ref{tab:model_pams_FPGA}.
The active policy model is continuously updated during training by using randomly selected experiences from the memory buffer. 
At each training step the weights of the target model $\theta_\mathrm{target}$ are incrementally updated to reflect the weights of the active policy model $\theta_\mathrm{policy}$,
\begin{equation}
\theta_\mathrm{target} \mapsto \theta_\mathrm{target}(1-\tau) + \tau  \theta_\mathrm{policy},
\end{equation}
where we set $\tau=0.5$~\cite{ddpg}.

\begin{figure}
\centering
\includegraphics[width=0.48\textwidth]{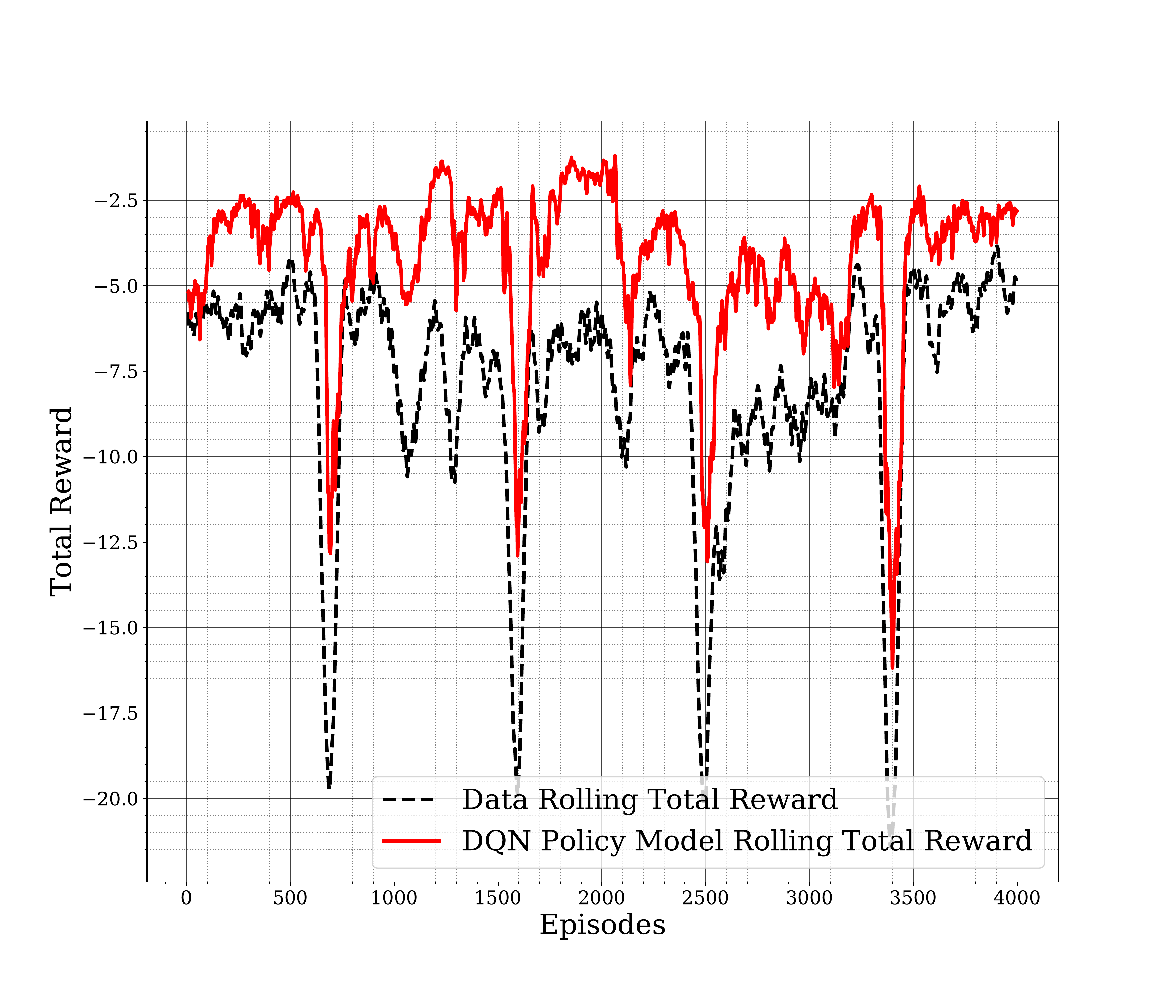}
\includegraphics[width=0.48\textwidth]{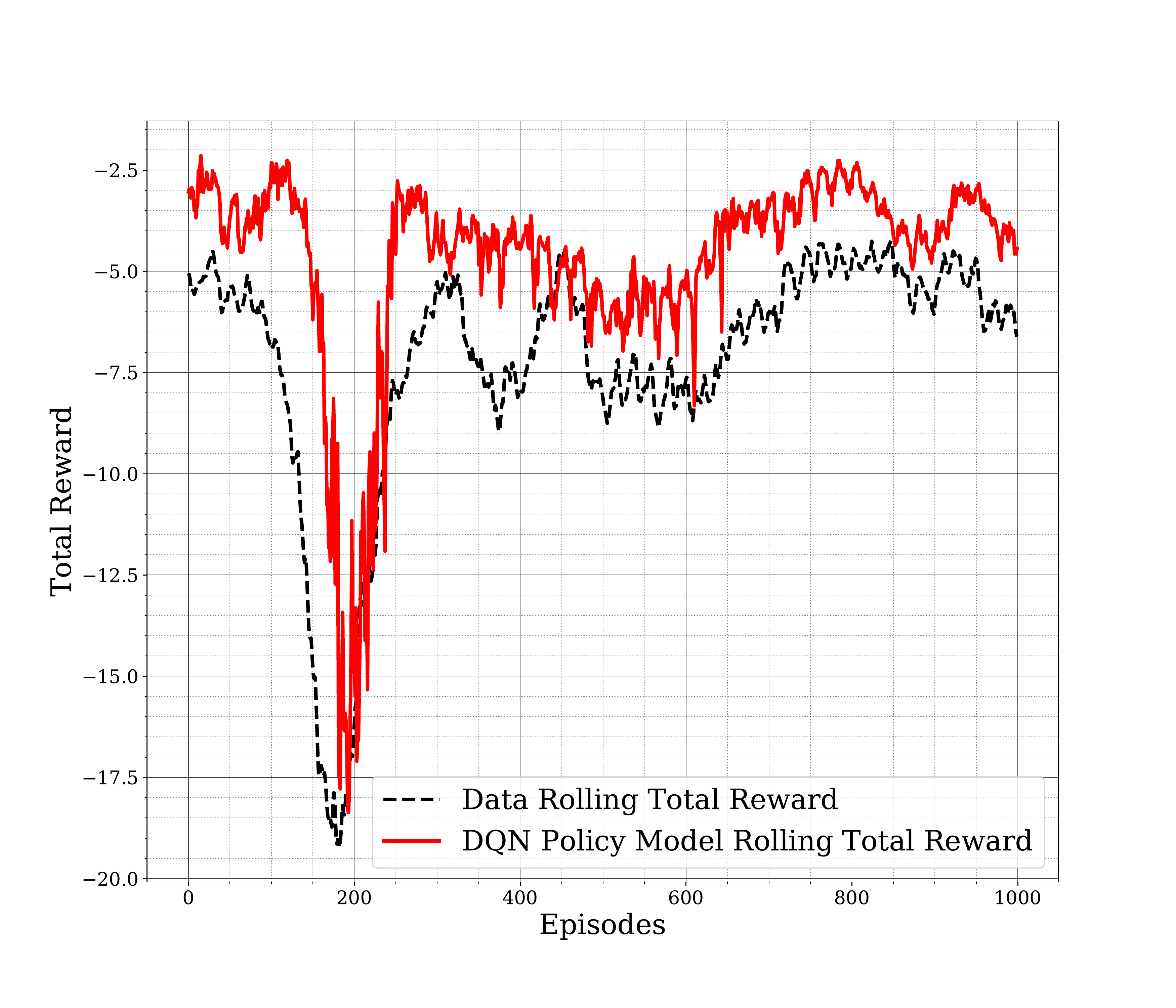}
\caption{Top: Total rolling reward per episode versus number of training episodes for the DQN MLP algorithm (solid red line) at top. During training, the 10-episode rolling window determines the first entry at the 10th episode on the plot. Bottom: The corresponding testing results (without a rolling average) are shown below.}
\label{fig:single-rl}
\end{figure}

The result of the DQN MLP training, in terms of a rolling average of the total reward over 10 episodes versus the number of episodes, is shown in Fig.~\ref{fig:single-rl} (top).
Here we show the results of consecutive batch initialization within a restricted region of 4,000 consecutive samples.
These samples fell within the first 250,000 samples that the surrogate model was trained on.
This was done to ensure that the environment would still reliably respond to the action.
Note that in the training, there is additional randomness introduced due to the epsilon-greedy approach.

Additionally, in Fig.~\ref{fig:single-rl} (bottom), we display the results of testing (no rolling average taken) versus the number of episodes.
Likewise, the testing start points were generated via consecutive batch initialization within an orthogonal region of 1,000 samples that still fell within the region the surrogate model was trained on.

For both the training and testing reward plots, the current controller reward, the black dashed line, was determined using the data by summing the values of $R_{t} = -|\BIMINER(t)|$ throughout each 50 step episode.
The DQN MLP controller reward, the solid red line, shows improvement over the current system by approximately a factor of 2 in both the training and testing sets.
The downward reward spikes in both the training and testing correspond to occurrences of a relatively rare but regular operation: the 6\,s resonant extraction from the Main Injector at 120\,GeV, sending beam through a long beam transport line, which requires certain large power supplies to operate at high current for the duration, strongly influencing the electrical environment of the GMPS. 
The RL policy is expected to learn to treat this appropriately once trained on more data. 
As is evident here, only experiencing these events four times is not enough for the RL policy to perfectly accommodate this circumstance.
Nevertheless the RL policy outperforms the current PID implementation in this context. 
It would be interesting to see the corresponding improvement to injection losses and Booster efficiency for a controller regulating GMPS with this RL policy, because these improvements are not thought to be linearly related to the size of the regulation error.

Additionally, we quantified the step by step difference between the RL agent's actions and the current PID's historical actions (defined as the change in \BVIMIN) over the same data during the test set.
The distribution of differences in action is approximately normal and has mean $0.0006 \pm 5.769 \times 10^{-5}$ and standard deviation $0.0129$, giving us confidence that the learned RL approach is reasonable.
We expect that future development work focused on finetuning the discretization of the action space will decrease this standard deviation.

\section{Implementation in Fast Electronics}
\label{sec:implfast}

Fast GMPS control electronics are required to collect information from the Booster environment, decide whether to apply a corrective action, and distribute the corresponding control signal, all with the low latency requirement set by the Booster's 15~Hz cycle. 
An FPGA is a natural choice to implement the corresponding circuit, accommodating latencies far below those achievable with a CPU or GPU while allowing reconfigurability impossible in a custom application-specific integrated circuit (ASIC) solution.
The DQN MLP GMPS regulation model proposed in Sec~\ref{sec:rl} requires an efficient, but adjustable, implementation of NN algorithms, strongly suggesting an FPGA-based implementation.
As a preliminary step, we take the offline-trained DQN MLP with weights fixed and deploy it in an FPGA.

The following subsections review the computational steps required for a single NN inference (\S\ref{sec:implfast:nn});
describe the basic elements of an FPGA and how a deep NN calculation can be efficiently mapped to a corresponding circuit (\S\ref{sec:implfast:fpga});
present an implementation of the DQN MLP described above in Sec~\ref{sec:methods} and the impact of various design choices (\S\ref{sec:implfast:impl});
and lastly discuss possible extensions of the implementation to accommodate more complex algorithms which are of interest (\S\ref{sec:implfast:future}).

\subsection{Elements of NN Inference}
\label{sec:implfast:nn}

The structure of an MLP is a series of alternating linear and nonlinear transformations (layers), with the \emph{i}th layer mapping a set of inputs $x_i$ (features) to a discrete list of outputs $y_i$.
In the present application, the features may include any measurements of the GMPS environment, such as digitized traces from the reference magnet system, line voltage frequency, and equipment gallery temperature.
For the DQN MLP, the outputs $y_i$ are scores associated to a discrete set of possible actions, with the highest-scoring action being the one taken by the controller.
An MLP layer $f$ yielding $m$ outputs may be written in terms of its action on a set of inputs $\{x_i\}_{i=1,\ldots,n}$ as
\begin{equation}
\label{eq:nnLayer}
f: x_i \to \sigma \left( \sum_{i} w_{ij} x_i + b_j \right),
\end{equation}
where $w_{ij}$ (the $n\times m$ weight matrix) and $b_j$ (the $m$-dimensional bias vector) are configurable parameters of the linear translation and $\sigma$ is an $m$-to-$m$ nonlinear activation function.
For each layer, the activation function is prescribed as a part of the model architecture while optimal values for the weights and biases are found through a training procedure.
The DQN MLP utilizes the linear (identity) $\mathrm{ReLU}(x_i) = \max(x_i,0)$ activation functions.
The complete, $k$-layer NN is specified by an ordered composition of layers $y = f^{(1)}f^{(2)}\ldots f^{(k)}(x)$.
While the input and output dimensions are fixed by the set of features and actions, the dimensionality of intermediate layers is arbitrary.
Table~\ref{tab:model_pams_FPGA} describes the architecture of the DQN MLP, in addition to the number of configurable parameters and total multiply-and-accumulate (MAC) operations required.

\begin{table}[!t]
\renewcommand{\arraystretch}{1.3}
\caption{Implemented DQN MLP model architecture. The first NN layer receives five input values.}
\centering
\begin{ruledtabular}
\begin{tabular}{r|c|c|c|c}
Layer & Outputs & Activation & Parameters & MACs \\
\hline
1 & 128 & ReLU & 768 & 640\\
2 & 128 & ReLU & 16\,512 & 16\,384\\
3 & 128 & ReLU & 16\,512 & 16\,384\\
4 & 7 & Linear & 903 & 896 \\
\hline
Total & \NA & \NA & 34\,695 & 34\,304
\end{tabular}
\end{ruledtabular}
\label{tab:model_pams_FPGA}
\end{table}

\subsection{NN Inference on FPGAs}
\label{sec:implfast:fpga}
An FPGA consists of an array of logic gates that may be programmed to emulate any circuit (up to the physical resource constraints of the specific hardware device).
This allows FPGA designs to profit from the many advantages of custom ASICs including massive parallelization and low power consumption while maintaining reconfigurability.
However, a significant advantage of the FPGA architecture lies in the fact that it is not simply a homogeneous fabric of low-level gates (e.g. NAND gates).
Rather, modern FPGAs are heterogeneous structures including more complex logical blocks, each specialized for a dedicated task, repeated many times.
In this way, FPGA designs can simultaneously exploit both the flexibility of a programmable architecture and the performance of a dedicated printed circuit.

An efficient implementation of the NN model in firmware requires a design that exploits the FPGA's specialized computational units to perform each step of the NN calculation.
Digital signal processor (DSP) slices are flexible circuits for addition, multiplication, wide-array bitwise logical operations, and more. DSPs may be further chained to accommodate more complex operations.
In the Intel Arria 10 FPGA, to be deployed in the GMPS control system, DSP blocks may be configured to multiply and accumulate fixed-point numbers up to 27 bits, providing a solution for the linear component of Equation~\ref{eq:nnLayer}.
The affine map from $m$ to $n$ dimensions requires $mn$ scalar multiplications and sums that, in a fully parallelized design, may be accomplished with $mn$ cascading DSPs.
To evaluate an arbitrarily complex activation function in FPGAs, it is more efficient to store a pre-computed table of values than to re-calculate the function many times per inference.
This may be accomplished using block RAM (BRAM), embedded memory that is configurable for read/write access.
BRAMs are available in segments of 20\,kb in the Arria 10 to store, for example, a bank of 1024 function values at 20-bit precision.
Registers are groups of flip flops used to record temporary numerical values or internal states, and to facilitate signal routing across the major computational blocks of the design.
Finally, ALMs are lightweight, configurable modules of combinational logic elements, used throughout designs for basic operations such as simple arithmetic and logical operations. 

\subsection{Implementation of the GMPS Regulator Model}
\label{sec:implfast:impl}

The GMPS regulator model described in Sec~\ref{sec:methods} must be converted to firmware in a manner that takes full advantage of the FPGA's architectural features described in Sec~\ref{sec:implfast:fpga}.
This is accomplished through the translation of the \textsc{Keras} description of the NN function into high-level synthesis (HLS) code using the \hlsfml~\cite{hls4ml} toolkit, whose functionality has been recently extended to Intel FPGAs~\cite{intelpaper}.
The HLS design is converted to firmware using \textsc{Intel Quartus}~\cite{quartus}. 
The use of \hlsfml brings the significant advantage of enabling a fast development cycle from model prototyping to implementation in firmware.
Thus, the present work has focused not only on achieving an optimal design for the benchmark ML algorithm proposed in Sec~\ref{sec:rl} but also on more generic design-space exploration.
Establishing scalable strategies such as FPGA implementation is critical for scaling up to more complex ML models that will inevitably become necessary as larger data sets allow for increasingly nuanced treatments of the control problem.

The conversion of the \textsc{Keras} model to firmware requires a number of design choices.
Chief among these are the numerical precision to which the calculation is carried out and the degree of parallelism incorporated into the design.
Fixed-point values with a specified number of total and integer bits are used to represent model inputs, weights, and all intermediate results of the calculation.
To determine the range of values to encode in the fixed-point representation, the number of integer bits is set to be at least as large as the maximum weight value.
The number of total bits, which sets the number of bits used to encode the fractional component of the weight (once the integer part is specified), is set to minimize the impact of quantization.

\begin{figure}
\centering
\includegraphics[width=0.48\textwidth]{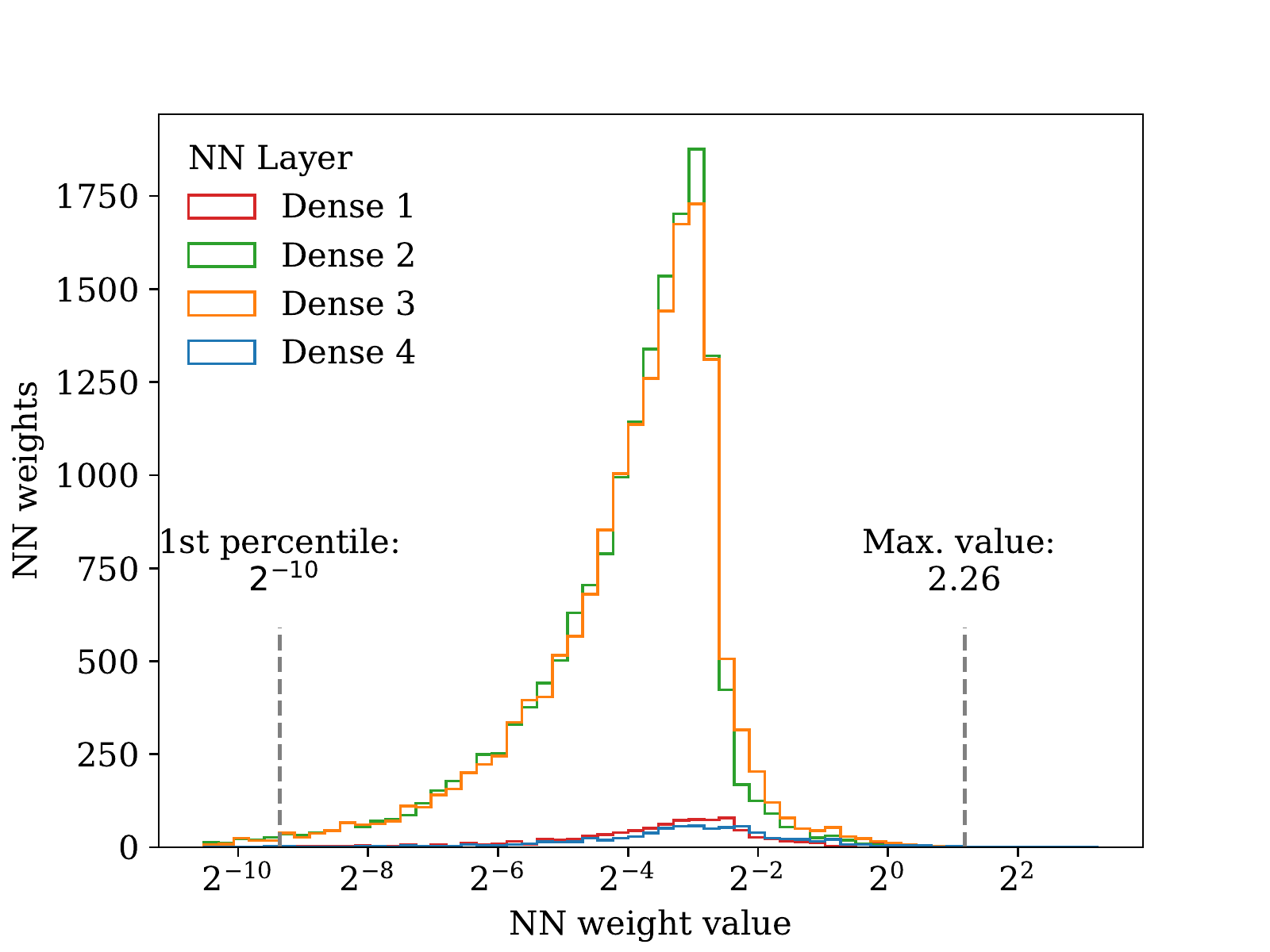}
\caption{Separate histograms display the magnitude of the floating-point weights obtained for each many-to-many (``dense'') layer after training the DQN MLP, in bins of logarithmically-varying width.  
Over 99\% of weights are found to have absolute value greater than $1/2^{10}$, with a maximum value of 2.26.}
\label{fig:nn_weights}
\end{figure}

\begin{figure}
\centering
\includegraphics[width=0.48\textwidth]{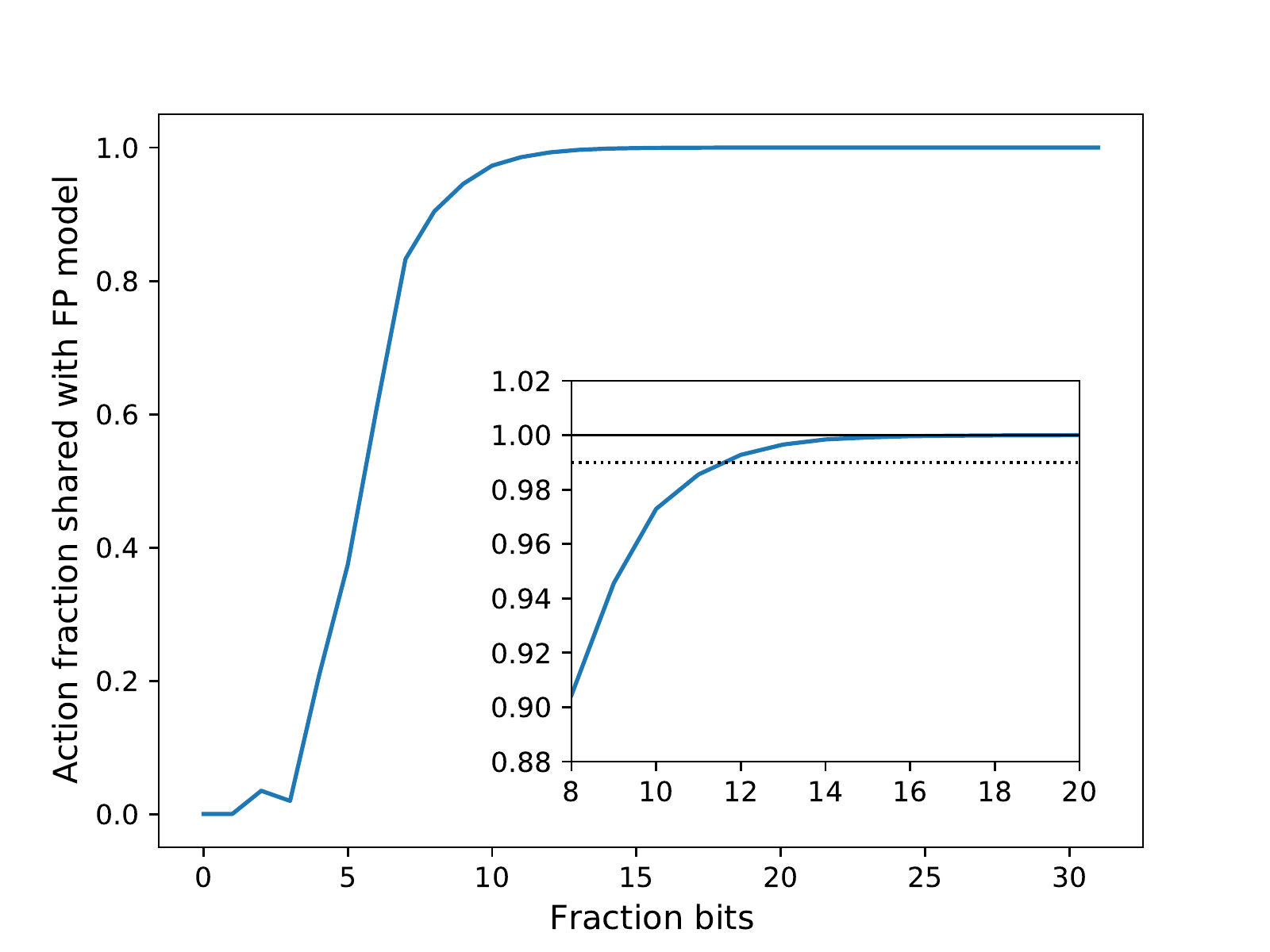}
\caption{The fraction of decisions that the quantized NN implementation shares with the floating-point calculation across a set of representative input Booster data is shown as a function of the fixed-point precision.
Here the number of bits to encode the integer part is fixed to five plus a sign bit, while the number of bits encoding the fractional part is varied.
The inset shows the same measurements, highlighting the region where the shared action fraction is over 90\%.
At very low precision, statistical fluctuations are observed that depend on the specific model weights and rounding conventions.}
\label{fig:nn_accuracy}
\end{figure}

Figure~\ref{fig:nn_weights} displays a histogram of the weight values for the trained DQN MLP model.
The total number of bits retained for each weight is selected by comparing the floating-point model inference with that of the fixed-point model for a range of bit widths, scanning over a representative sample of input test data.
Figure~\ref{fig:nn_accuracy} shows that using 14 bits to encode the fractional component of all operands is sufficient to replicate the decision taken by the floating-point model for over 99.5\% of the test data.
This performance comparison of fixed and floating-point models shows that while nine fractional bits are sufficiently precise to represent the weights, additional precision in the representation of the intermediate sums in the NN calculation is necessary to achieve full performance.
Following this, comparisons were performed using a 20-bit representation for all internal fixed-point parameters (5 integer bits, 14 fractional bits, and a sign bit) where not explicitly varied.

The degree of design parallelization can be motivated by the number of operations required for each NN inference in comparison with the total available resources on the target FPGA.
One constraint comes from the MAC operations that are efficiently computed in a single clock cycle using dedicated DSP slices for each operation.
As discussed in Sec~\ref{sec:implfast:nn}, the MLP agent requires 34\,304 MACs per inference, compared to the 1518 DSP slices available in the Arria 10 FPGA.
The approach taken to address this is to assign a reuse factor to each NN layer, specifying the number of operations each physical MAC unit may contribute to the set of necessary computations required by that layer.
Larger reuse factors will result in a design utilizing fewer FPGA resources at the expense of longer inference latency.
For simplicity a single, model-level reuse factor is considered in the following, where the per-layer reuse factor is given by the greatest common divisor of the reuse factor and the product of input and output multiplicity.

\begin{figure}
\centering
\includegraphics[width=0.48\textwidth]{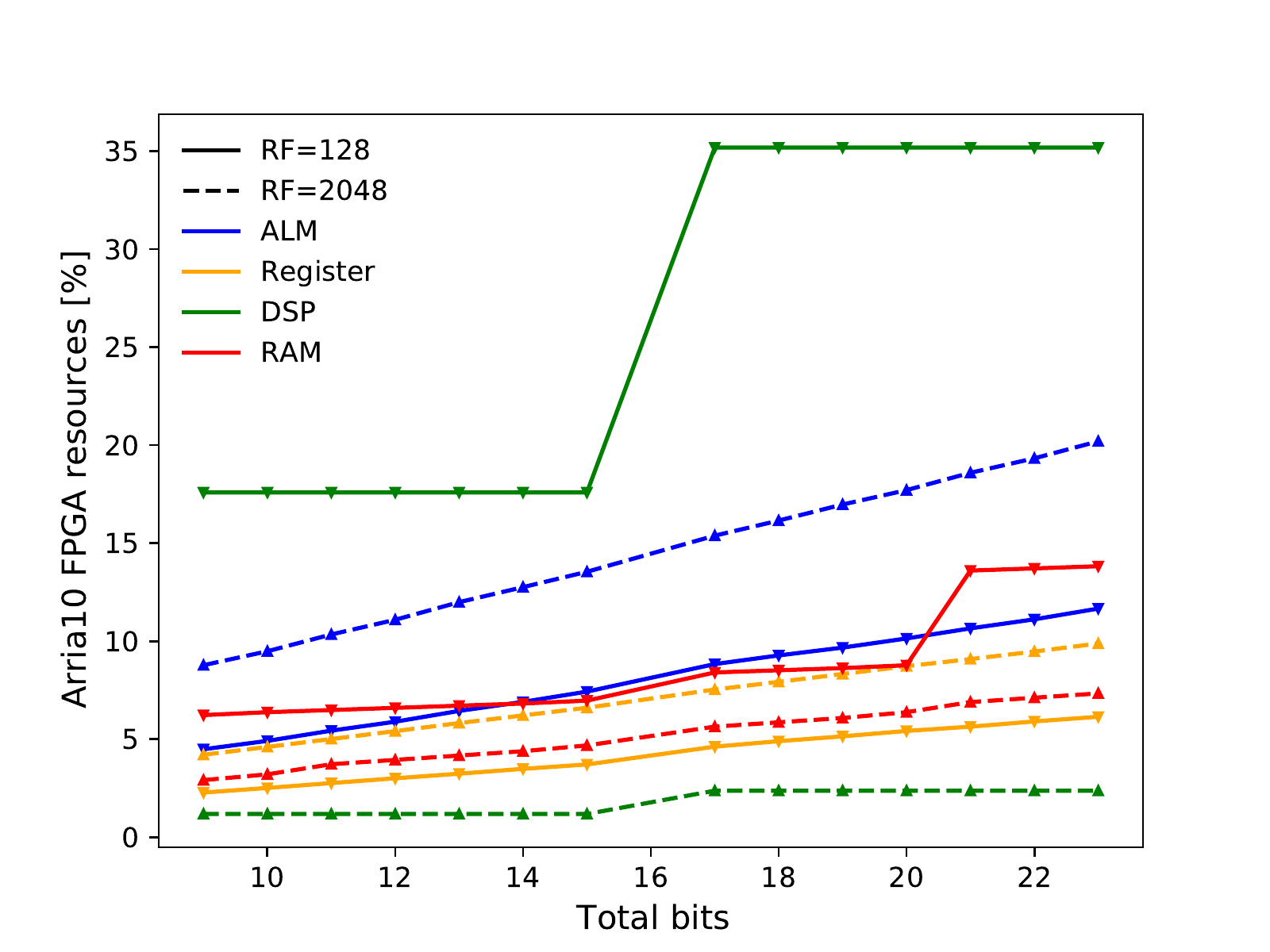}
\caption{FPGA resources required for the implementation of the DQN MLP are shown as a function of the fixed-point precision utilized for internal NN operations.  
All resources are normalized to the total available in a benchmark Arria 10 device (see Table~\ref{tab:resources}).
Results are shown for implementations with reuse factors of 128 (solid lines) and 2048 (dashed lines).}
\label{fig:resource_scan}
\end{figure}

Figure~\ref{fig:resource_scan} demonstrates how the resources required to implement the NN algorithm are affected by the precision to which internal calculations are carried out, and the degree of parallelization specified by the selected reuse factor.
In general, the required low-level resources such as ALMs and registers scale linearly with precision to accommodate the widths of increasing data paths.
Conversely, a single DSP slice can accommodate a range of operand bit-widths, up to the limit of the design specification at which point a second DSP must be used per calculation.
In the case of reuse factor of 128, the largest burden on FPGA resources comes from the required DSPs (either 4 or 8\% of the Arria 10 total, depending on the necessary precision), while ALMs are the limiting factor (3--8\%) in the case of reuse factor of 1568, where the design parallelization is at most a factor of four in each NN layer.
While inference latency depends on the degree of parallelism directly through the reuse factor, it is essentially invariant under changes to the operand precision.

Table~\ref{tab:resources} compares several implementations for constant precision (20 total bits) and various reuse factors.
In general the algorithm latency increases as a function of increasing reuse factor while the numbers of DSPs and BRAMs required are inversely proportional to the reuse factor.
Variations in the required registers and ALMs are generally not significant by comparison.
These results demonstrate a range of feasible firmware implementations of the algorithm that fit comfortably within the available resources of the GMPS control board and 1/15\,sec (66.7\,ms) latency budget.
The ability to tune resource usage provides significant flexibility to accommodate future scenarios where the NN algorithm may significantly grow in complexity and, further, must coexist on a single FPGA with additional control logic that may present inflexible resource constraints of its own.

\begin{table}[!t]
\renewcommand{\arraystretch}{1.3}
\caption{
The required FPGA resources and corresponding latency for the NN algorithm are shown for three possible implementations corresponding to various reuse factors.
In addition to design parameters, the maximum available resources are shown for an Intel Arria 10 benchmark FPGA. 
Memory logic array blocks (MLABs) are configured from ten ALMs and hence no device maximum is shown.
}
\centering
\begin{ruledtabular}
\begin{tabular}{r||c|c|c|c|c||c}
reuse factor & DSP & BRAM & MLAB & ALM & Register & Latency \\
\hline
128  & 534 & 238 & 672 & 43.3\,k & 92.6\,k & 3.9\,$\mu$s  \\
256 & 274 & 231 & 642 & 48.9\,k & 112.3\,k & 7.3\,$\mu$s \\
512 &  144 &  195 & 7467 & 152.6\,k & 252.0\,k & 13.6\,$\mu$s \\
1024 &  68 &  171 & 4088 & 111.7\,k & 202.4\,k & 24.6\,$\mu$s \\
2048 &  36 &  173 & 1960 & 75.6\,k & 149.1\,k & 39.3\,$\mu$s \\
\hline
Available & 1518 & 2713 & \NA & 427\,k &1.7\,M & \NA \\
\end{tabular}
\end{ruledtabular}
\label{tab:resources}
\end{table}

\subsection{Extensions to More Complex Algorithms}
\label{sec:implfast:future}

Up to this point, the discussion of the hardware implementation has centered around the three-hidden-layer MLP architecture found to be performant for the GMPS control problem in the context of RL studies described in Sec~\ref{sec:methods}.
However, the conclusions of the studies described above may be extended to more complex NN algorithms providing improved GMPS performance in tandem with the experience gained through future data-taking campaigns.

The simplest extension to the single MLP solution, well-motivated in the context of RL studies, is to run inference with an ensemble of multiple copies of the network in parallel on the FPGA, to improve robustness of performance.
Each NN may be programmed with a unique set of weights, allowing for disagreement among the models, where additional voter logic determines the final action to be taken by the control system.
This is straightforward to achieve for models with similar complexity to the one studied in Sec~\ref{sec:implfast:impl}.
Achieving designs that consume $\leq$6\% of all available resources suggests that an ensemble of $\mathcal{O}(10)$ models is feasible.

Alternatively, instead of an ensemble of relatively simple models, more complex networks can be pursued.
The MLP architecture studied can be extended to additional layers and larger numbers of nodes per layer maintaining an acceptable footprint through corresponding adjustment of the reuse factor.
The theoretical scaling behavior was shown in the calculations of Sec~\ref{sec:implfast:nn} and observed in the implementation using Quartus HLS.
As an illustrative example, one could consider a refinement of the baseline architecture where the number of nodes per layer is uniformly increased by a scaling factor $s$. 
In this case, the number of required multipliers may be kept constant by simultaneously increasing the reuse factor by a factor of $s^2$, at the expense of a small corresponding increase in algorithm latency.
This strategy would allow more powerful solutions with similar footprint to take advantage of the full latency budget of $\approx66$\,ms.
More sophisticated architectures such as convolutional and recurrent NNs may also be considered, taking advantage of their representations as compositions of multiple dense sub-layers.
A detailed study of such possibilities is left to future work.

\section{Summary and Outlook}
\label{sec:summary}
 
In this paper, we have described a method for controlling the gradient magnet power supply (GMPS), an important subsystem of the Fermilab Booster accelerator, using machine learning models and demonstrated the feasibility of embedding such a model on a field-programmable gate array (FPGA) for a high-uptime, low-latency implementation.
We first developed a surrogate LSTM model, based on a recurrent neural network, to reproduce the behaviors of the real GMPS system in the context of the accelerator complex, establishing a safe environment for training reinforcement learning algorithms.
Within this environment, we trained a deep $Q$-network, based on a multilayer perceptron, to choose an optimal action (adjustment of one control knob) to maximize the long-term reward, taken from the negative absolute value of the regulation error (difference between the set and observed values of the minimum GMPS current).
We found this surrogate-trained network achieved a factor of 2 improvement over the existing controller in terms of the achieved rewards. 
Finally, we implemented this network on an Intel Arria 10 FPGA and found it reproduces the CPU-based model, consumes less than 6\% of the total FPGA resources, and executes with a latency as low as 2.8\,$\mu$s, which bodes well for future extensions.

Real-time and operations-hardened solutions will be critical for deploying this technology in an accelerator control context, but we believe a large number of other application spaces will be able to benefit from reinforcement learning on embedded systems.
Surrogate models appear promising for supplying the large training data volumes required by reinforcement learning agents. 
This is particularly important for accelerator facilities where large-scale simulations of the entire complex are absent.
Although many open questions remain, this proof-of-principle provides confidence to test our proposed concept on ``live'' hardware.
The next steps of this work, including mechanisms for online training and model updates for systems operating with a running accelerator, will be the subject of a future report.
 
In general, the future for machine learning algorithms in accelerator control is bright.
The proliferation of shared tools and open datasets like the one developed for and used in this paper will doubtlessly enable rapid progress.
We note that Ref.~\cite{cernrl} also adopted the \textsc{OpenAI Gym}~\cite{openai_gym} as a programming interface for training reinforcement learning agents for use in an accelerator complex.
Adoption of common tools will make it easy for researchers in this space to share code and especially to share access to datasets.
This should allow multi-institution collaboration to prosper and greatly enhance the pace of progress in the field of artificial intelligence for accelerator applications.
 
\begin{acknowledgments}

We would like to thank Jonathan Edelen and Jan Strube for useful discussions and a careful reading of the manuscript.

This document was prepared using the resources of the Fermi National Accelerator Laboratory (Fermilab), a U.S. Department of Energy (DOE), Office of Science, HEP User Facility. 
Fermilab is managed by Fermi Research Alliance, LLC (FRA), acting under Contract No. DE-AC02-07CH11359.
This research was sponsored by the Fermilab Laboratory Directed Research and
Development Program under Project ID FNAL-LDRD-2019-027: ``Accelerator Control with Artificial Intelligence'' and partially funded by the National Science Foundation (NSF) Mathematical Sciences Graduate Internship Program (MSGI).

Pacific Northwest National Laboratory (PNNL) is a multi-program national laboratory operated by Battelle for DOE under Contract No. DE-AC05-76RL01830. 
This work at PNNL was supported by the DOE Office of High Energy Physics.

J.~M.~D. is supported by DOE Office of Science, Office of High Energy Physics Early Career Research program under Award No. DE-SC0021187. 
R.~K. was supported by the NSF MSGI and the NSF Graduate Research Fellowship Program under Grant No. DGE-1644869.

We acknowledge the Fast Machine Learning collective as an open community of multi-domain experts and collaborators. 
This community was important for the development of the FPGA implementation for this project.

\end{acknowledgments}

\bibliography{references}

\end{document}